\documentclass[preprint,showpacs,superscriptaddress,preprintnumbers,amsmath,amssymb]{revtex4}

\usepackage[pdftex]{graphicx}
\usepackage{dcolumn}
\usepackage{bm}
\usepackage[thinspace]{SIunits}

\begin{document}


\title{Enhanced photoluminescence emission from
two-dimensional silicon photonic crystal nanocavities}

\author{N. Hauke}
\email{hauke@wsi.tum.de}
\affiliation{Walter Schottky Institut, Technische Universit\"at M\"unchen, Am Coulombwall 3, D-85748 Garching, Germany}%
\author{T. Zabel}
\affiliation{Walter Schottky Institut, Technische Universit\"at M\"unchen, Am Coulombwall 3, D-85748 Garching, Germany}%
\author{K. M\"uller}
\affiliation{Walter Schottky Institut, Technische Universit\"at M\"unchen, Am Coulombwall 3, D-85748 Garching, Germany}%
\author{M. Kaniber}
\affiliation{Walter Schottky Institut, Technische Universit\"at M\"unchen, Am Coulombwall 3, D-85748 Garching, Germany}%
\author{A. Laucht}
\affiliation{Walter Schottky Institut, Technische Universit\"at M\"unchen, Am Coulombwall 3, D-85748 Garching, Germany}%
\author{D. Bougeard}
\affiliation{Walter Schottky Institut, Technische Universit\"at M\"unchen, Am Coulombwall 3, D-85748 Garching, Germany}%
\author{G. Abstreiter}
\affiliation{Walter Schottky Institut, Technische Universit\"at M\"unchen, Am Coulombwall 3, D-85748 Garching, Germany}%
\author{J. J. Finley}
\affiliation{Walter Schottky Institut, Technische Universit\"at M\"unchen, Am Coulombwall 3, D-85748 Garching, Germany}%
\author{Y. Arakawa}
\affiliation{Institute for Nano Quantum Information Electronics, Institute of Industrial Science, The University of
Tokyo, 4-6-1 Komaba, Meguro, Tokyo 153-8505, Japan}

\date{\today}

\begin{abstract}
We present a temperature dependent photoluminescence study of silicon optical nanocavities formed by introducing point defects into two-dimensional photonic crystals. In addition to the prominent TO phonon assisted transition from crystalline silicon at $\sim1.10$ eV we observe a broad defect band luminescence from $\sim1.05-1.09$ eV. Spatially resolved spectroscopy demonstrates that this defect band is present only in the region where air-holes have been etched during the fabrication process.  Detectable emission from the cavity mode persists up to room-temperature, in strong contrast the background emission vanishes for $T\geq150$~K.  An Ahrrenius type analysis of the temperature dependence of the luminescence signal recorded either in-resonance with the cavity mode, or weakly detuned, suggests that the higher temperature stability may arise from an enhanced internal quantum efficiency due to the Purcell-effect.
\end{abstract}

\maketitle
Due to its indirect electronic bandgap, crystalline silicon is rarely used as the active emitter in semiconductor optics \cite{Pavesi04}. Interband light emission is predominantly a phonon-assisted process and silicon has, therefore, a very poor internal quantum efficiency. However, the development of efficient silicon based light emitters would pave the way toward CMOS compatible monolithic optical interconnects and, therefore, signal processing speeds much higher than currently provided by silicon micro-electronics \cite{Jalali07,Jalali06,Soref06,Izhaky06,Kimerling00}.

Recently, enhancement of the photoluminescence (PL) intensity from crystalline silicon at room-temperature has been observed using two-dimensional (2D) silicon photonic crystals (PhCs) with photonic point defect nanocavities \cite{Iwamoto07,Fujita08}. Such defect PhC nanocavities modify the spatial emission profile of light; due to the in-plane photonic band gap a much larger fraction of luminescence is emitted perpendicular to the slab than parallel to it \cite{Kaniber07}. In addition, the high $Q$-factors that are attainable using PhC nanocavities \cite{Akahane05,Song05,Asano06} together with their small mode volumes may lead to an enhancement of the internal quantum efficiency of the active material due to an increase of the radiative emission rate via the Purcell-effect \cite{Purcell46}. Other work demonstrates enhanced PL emission from internal light emitters embedded in photonic crystal slabs \cite{Kurdi08,Xia09}.

In this letter, we present a detailed investigation of the spectrum and temperature stability of the PL emission from crystalline silicon PhC nanocavities. Analysis of our results indicates that both the phonon satellites of the interband silicon emission and surface defect states are responsible for the luminescence of PhC cavity modes. Most interestingly, we can detect luminescence from the cavity mode up to room temperature whilst the background emission intensity rapidly reduces below our detection threshold for $T\geq150$~K.  An Arrhenius type analysis of the temperature dependent data, either in resonance with the cavity mode or spectrally detuned from it, indicates that the local radiative emission rate is enhanced in resonance via the Purcell-effect.

\begin{figure}[t!]
\includegraphics[width=0.75\columnwidth]{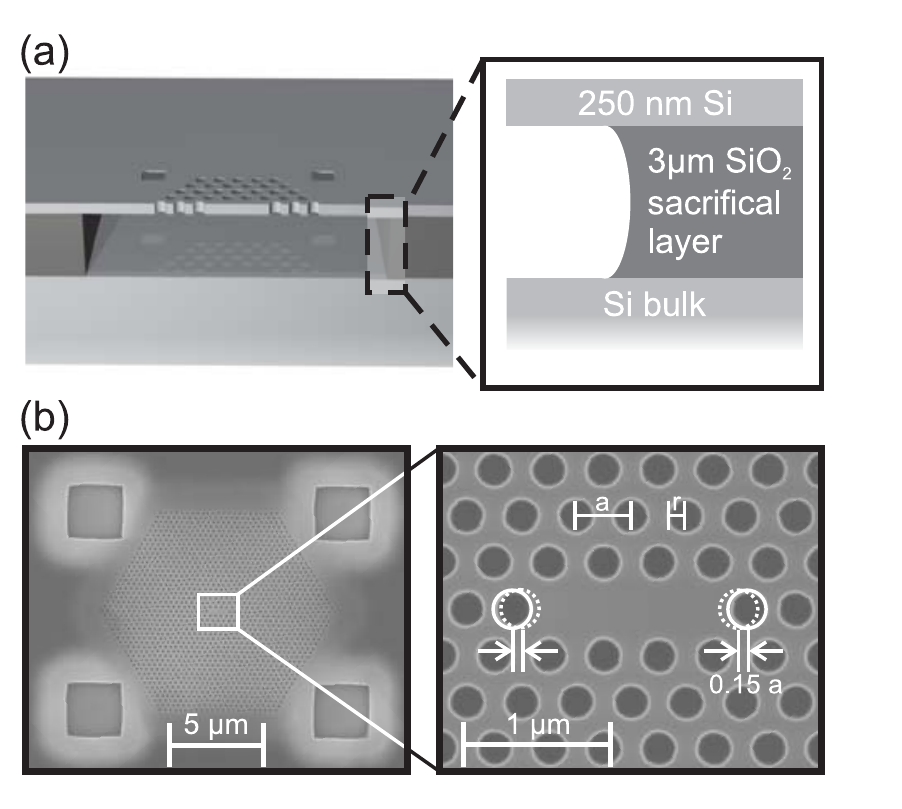}
\caption{\label{figure1}
(a) Left panel: schematic cross-sectional representation of the photonic crystal nanocavity structures investigated. Right panel: layer sequence in the active region.
(b) Left panel: SEM image showing a photonic crystal from top. Right panel: zoom-in to the modified L3 defect where the outer holes are shifted by 0.15 lattice constants.}
\end{figure}

The samples investigated were fabricated from silicon-on-insulator (SOI) wafers with a $d = \unit{250}{\nano\metre}$ thick active silicon layer on top of a $\unit{3}{\micro\metre}$ thick layer of SiO$_2$.
A 2D PhC is patterned into the upper silicon layer as illustrated in \mbox{fig. \ref{figure1} (a)}. We defined a triangular lattice of air holes with a period of $a = \unit{275}{\nano\metre}$ in the silicon slab as shown by the scanning electron microscope (SEM) image in the left panel of \mbox{fig. \ref{figure1} (b)}.
%
This was done using electron-beam-lithography and subsequent SF$_6$/C$_4$F$_8$ reactive-ion-etching. These techniques allow us to control the radius $r$ of the air holes with a precision of $\pm\unit{2}{\nano\metre}$. As a final fabrication step the underlying SiO$_2$ is removed by hydrofluoric acid. Low mode-volume nanocavities ($V_{Mode} \approx 1 \cdot(\lambda / n)^3$) were realized by omitting three air-holes in a row and by shifting the lateral holes away from the cavity center by 0.15 lattice constants to form modified $L3$ cavities \cite{Akahane03}. A typical scanning electron micrograph (SEM) image of the samples investigated is shown in the right panel of \mbox{fig. \ref{figure1} (b)}.

Spatially-resolved optical measurements were performed using a micro-photoluminescence ($\mu$PL) spectroscopy setup. The sample was excited by a diode-pumped CW solid-state-laser emitting at $\lambda_{Laser} = \unit{532}{\nano\metre}$. We focussed the laser beam using a $50\times$ microscope objective ($NA=0.5$) to a spot size of $\approx \unit{1}{\micro\metre}$. The resulting PL signal was collected through the same objective and dispersed by a $\unit{0.32}{\metre}$ imaging monochromator equipped with a 600 lines/mm grating and a liquid nitrogen-cooled \text{InGaAs} linear diode array.

\begin{figure}[t!]
\includegraphics[width=0.75\columnwidth]{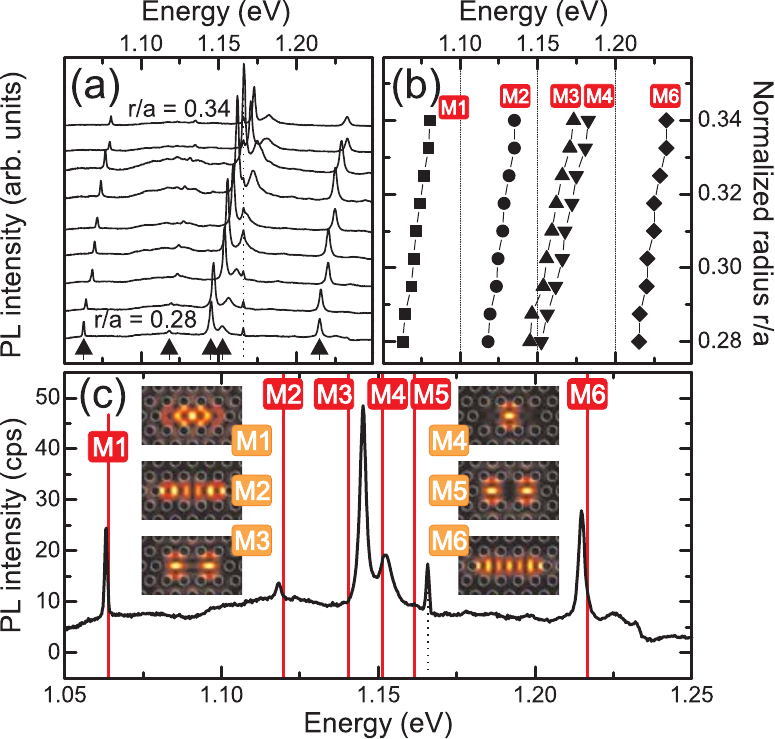}
\caption{\label{figure2}(a) Room-temperature PL spectra recorded from a series of L3 PhC cavities with different normalized air-hole radii ranging from $r/a=0.28$ (bottom) to $r/a=0.34$ (top).
(b) Detailed analysis of the five peaks (labeled M1 to M6, marked with arrows in (a)) shifting to higher energy with increasing $r/a$.
(c) Comparison between experimental PL data (black curve)  and simulated spectral mode positions (vertical red lines) of the six L3 modes for $r/a=0.28$. Insets show calculated electric field profiles of the eigenmodes. }
\end{figure}

In \mbox{fig. \ref{figure2} (a)} we present room-temperature $\mu$PL spectra recorded from a series of $L3$ nanocavities as the normalized radius $r/a$ is increased from 0.28 (bottom) to 0.34 (top).
The excitation power density used in these measurements was $\unit{550}{\kilo\watt\centi\rpsquare\metre}$. Using these measurement conditions the intensity of the emission from the unpatterned region of the device was below our detection sensitivity.  In strong contrast, each of the spectra recorded from the PhC nanocavities clearly reveals five distinct peaks (marked with arrows on fig. \ref{figure2} (a)) that shift to higher energy with increasing $r/a$ ratio \footnote{The fixed peak at $\unit{1.165}{\electronvolt}$ (marked with dashed line) arises from the $\unit{1064}{\nano\metre}$ emission of the laser.}. The extracted peak energies are plotted in \mbox{fig. \ref{figure2} (b)}, clearly demonstrating that all peaks (labeled M1 to M6) shift in a similar way to higher energies with increasing $r/a$, as previously reported in refs. \cite{Iwamoto07,Fujita08}. In \mbox{fig. \ref{figure2} (c)} we compare the measured PL spectrum  for $r/a = 0.28$ with the spectral positions of mode-emission obtained from photonic bandstructure simulations (vertical red lines). Using the geometric parameters extracted from SEM images, we obtain six eigenmodes from the simulation results. The insets in \mbox{fig. \ref{figure2} (c)} show the calculated electric field distributions that reveal three modes with longitudinally oriented field distributions (M1,M2,M6) and three modes with vertically oriented field distributions (M3,M4,M5). In full accordance with Iwamoto \emph{et al.} \cite{Iwamoto07}, we observe the strongest peak intensities for modes that have a high vertical extraction efficiency (M1,M3,M6). Simulation and experiment for the longitudinal modes are generally in very good agreement. A discrepancy is observed between simulation and experiment for M5, which is not observed in the spectra, probably due to a higher sensitivity of mode $Q$ and frequency to structural disorder.  This tends to lead to a low-Q and, thus, broad emission from M5. The comparison between simulation and experiment clearly shows that we observe PhC mode-emission in the spectral vicinity of the silicon interband luminescence and that we are able to clearly distinguish the cavity mode emission from the background.
%
%

\begin{figure}[t!]
\includegraphics[width=0.75\columnwidth]{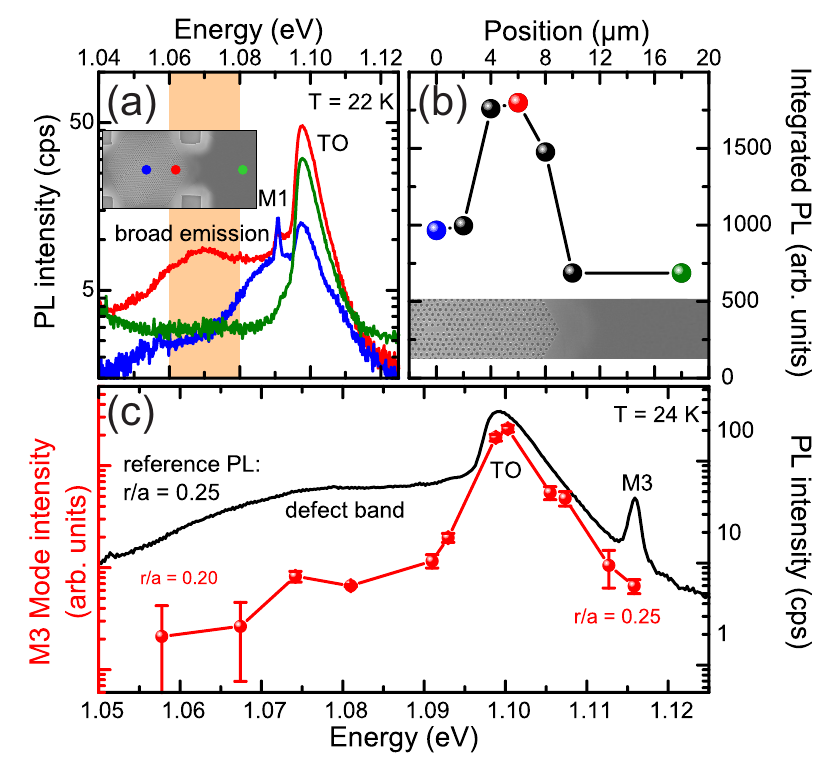}
\caption{\label{figure3}(a) PL at $T=\unit{22}{\kelvin}$ for three different detection positions (indicated by the inset) on the sample structure.
(b) Integrated PL of broad emission band at the low-energy side of the TO-replica (range highlighted orange in (a)) when scanning across the PhC. Inset shows correlation with etched air-hole structure.
(c) Red data: integrated mode intensity of M3 as a function of emission energy for $r/a$ ranging from 0.25 to 0.20. Black line: reference PL spectrum for $r/a=0.25$ at $T=\unit{24}{\kelvin}$.}
\end{figure}

In \mbox{fig. \ref{figure3} (a)} we compare low temperature ($T=\unit{22}{\kelvin}$) $\mu$PL spectra recorded close to the M1 mode emission from three different spatial positions as indicated on the inset.  The measurement positions were: on the nano-cavity (blue trace), on the PhC away from the cavity (red trace) and on the unpatterned material (green trace). The peak at $\approx \unit{1.10}{\electronvolt}$ arises from the TO-phonon replica of the interband emission from the silicon \cite{Hull99}.
%
In addition to the TO-line, we observe a broad emission band when detecting on the PhC that extends from $\sim1.05-1.09$~eV on the low energy side of the TO-replica . In bulk silicon one would not expect emission in this range at low temperatures, since all phonon replica are sharply defined \cite{Hull99}. We investigated the intensity of the broad emission band as a function of the excitation position on the sample.  Selected results of these measurements are presented in \mbox{fig. \ref{figure3} (b)} where we plot the integrated PL intensity from $\unit{1.06}{\electronvolt}$ to $\unit{1.08}{\electronvolt}$ (highlighted orange in (a)) for a spatial scan across the PhC (compare with fig. \ref{figure3} (a)-inset). Here, we clearly observe a direct correlation between PL intensity of the emission in the $1.06-1.08$~eV band and the number of air-holes in the probed region. This strongly indicates that the broad emission band arises from phonon-mediated recombination from surface defects on the sidewalls of the air holes created by the RIE etching process. A large number of surface defect states with different trapping energies would lead to the inhomogeneously broadened emission band on the low energy side of the TO-phonon replica. To investigate the origin of the emission in the PhC cavity modes we controllably shifted M3 through the TO-phonon line and the surface defect band by progressively decreasing the air-hole radius from $r/a=0.25$ to $r/a=0.20$ and extracted the intensity of the M3 emission. The results are summarized in \mbox{fig. \ref{figure3} (c)}, where we plot the integrated intensity of M3 as a function of peak position (red data points). For comparison, we plot the PL spectrum recorded from the $r/a=0.25$ sample (black line). The mode intensity clearly tracks the spectrum of the TO replica and follows also the intensity trend of the surface defect band. These observations indicate that the cavity modes are pumped predominantly via the phonon replica and also more weakly via the surface defects introduced by the fabrication process. This arises from the fact that the luminescent defects are close to the air-silicon interface, where the electric field amplitude of the eigenmodes as shown in \mbox{fig. \ref{figure2} (c)} are weak, leading to a weak coupling between mode and surface defects.

We continue to analyze the temperature stability of the PL spectrum recorded from the PhC cavity site. Typical temperature dependent data recorded from the M1 mode emission for a PhC with $r/a=0.34$ from \unit{24}{\kelvin} to room-temperature (\unit{295}{\kelvin}) is presented in \mbox{fig. \ref{figure4} (a)}.
We observe a clear and rapid reduction in the intensity of the TO-phonon replica, the M1 mode emission and the surface defect band, respectively. This is caused by an increasing importance of non-radiative decay channels, like Auger and free carrier recombination \cite{Dumke62}. For further quantitative analysis we describe the temperature dependent PL
intensity $I_{PL}(T)$ as a function of the external photon detection probability $P$, the excitation rate $R$ and the internal quantum efficiency $\eta_{int}(T)$:
\begin{figure}[t!]
\includegraphics[width=0.75\columnwidth]{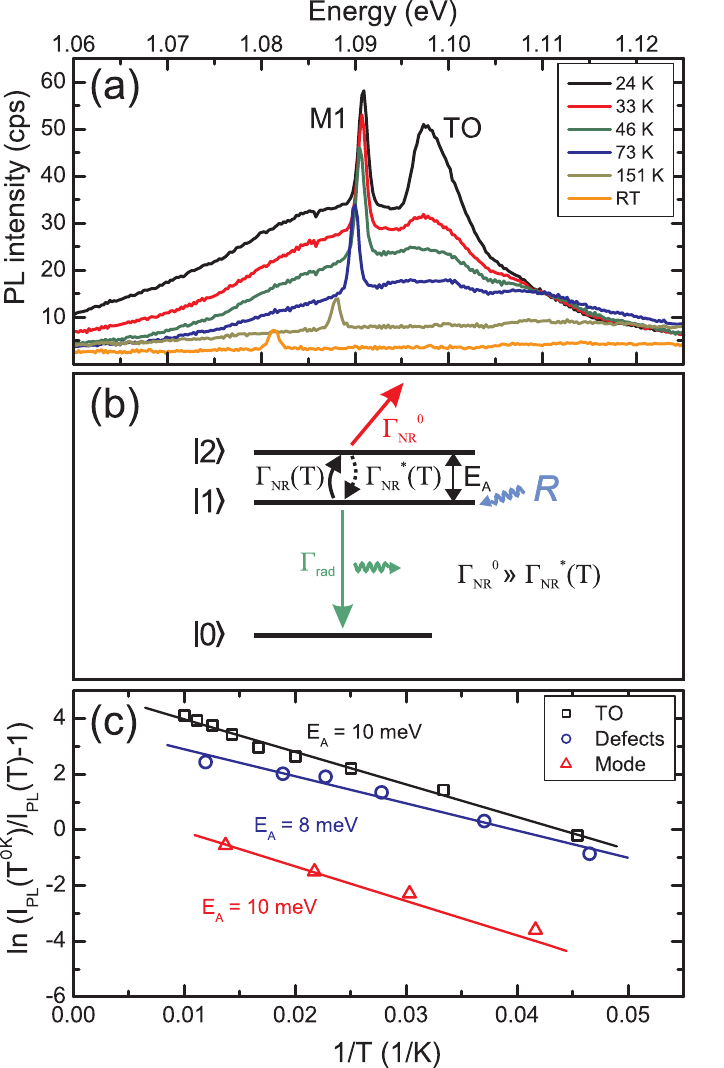}
\caption{\label{figure4}(a) PL spectra as a function of temperature for a PhC with $r/a=0.34$.
(b) Schematic level diagram illustrating the decay from state $|1\rangle$ via a radiative channel to state $|0\rangle$ and via a non-radiative channel over state $|2\rangle$.
(c) Arrhenius-type analysis for the TO-replica (black squares), defect band emission (blue circles) and M1 mode emission (red triangles).  The activation energies $E_A$ extracted from linear fits label the three curves}
\end{figure}
\begin{subequations}
    \begin{equation}
    I_{PL} (T) = P \cdot R \cdot \eta_{int}(T),
    \label{eq:1}
    \end{equation}
    \begin{equation}
    \text{with } \eta_{int}(T) =  \frac{\Gamma_{Rad}}{\Gamma_{Rad} + \Gamma_{NR}(T) }
    \label{eq:2}
    \end{equation}
    \begin{equation}
    \text{and } \Gamma_{NR}(T) = \Gamma_{NR}^0 \cdot \text{exp}\left(\frac{-E_A}{k_B T}\right),
    \label{eq:3}
    \end{equation}
\end{subequations}
In these equations $\Gamma_{Rad}$ is the interband radiative recombination rate and $\Gamma_{NR}(T)$ is the non-radiative recombination rate. The temperature-dependence of $\Gamma_{NR}(T)$ is described by distributing excitation amongst the two levels, $|2\rangle$ and $|1\rangle$ separated by $E_A$ according to Boltzmann statistics.  A schematic few level-diagram of the system is presented in \mbox{fig. \ref{figure4} (b)}. State $|1\rangle$ is pumped at an excitation rate $R$ and decays either via a radiative channel to $|0\rangle$ or non-radiatively over $|2\rangle$. \mbox{Eqn. \ref{eq:1}} to \mbox{eqn. \ref{eq:3}} are directly obtained from rate equations governing the steady state populations of $|1\rangle$ and $|2\rangle$ using the approximation $\Gamma_{NR}^0 \gg \Gamma_{NR}^{\ast}(T)$.  This condition ensures that the non-radiative channel dominates carrier recombination at elevated temperature as expected for silicon. By reformulating eqn. \ref{eq:1} we obtain
    \begin{equation}
    \text{ln}\left(  \frac{I_{PL}(T^{0K})}{I_{PL}(T)}-1 \right) = \frac{-E_A}{k_B T} - \text{ln}\left(  \frac{\Gamma_{rad}}{\Gamma_{NR}^0} \right),
    \label{eq:6}
    \end{equation}
where $I_{PL}(T^{0K})$ is the extrapolated PL intensity for \mbox{$T\rightarrow \unit{0}{\kelvin}$} \footnote{Equal to $I_{PL}(T)$ in absence of a non-radiative channel.}. In \mbox{fig. \ref{figure4} (c)} we plot \mbox{ln($I_{PL}(T^{0K})/I_{PL}(T) -1)$}, as a function of inverse temperature in an Arrhenius-type graph, according to eqn. \ref{eq:6}.  The analysis was performed by spectrally integrating over the TO-replica (black squares), the surface defect band PL from \unit{1.07}{\milli\electronvolt} to \unit{1.08}{\milli\electronvolt} (blue circles) and the M1 cavity mode emission (red triangles) \footnote{The intesity data for the TO-replica and defect band was obtained when detecting on the PhC region but \textit{away} from the cavity.}. For all datasets we observe a straight line as predicted by eqn. \ref{eq:6} supporting the vailidity of our analysis.  From the slopes of these curves we extract an activation energy $E_A$ of $\unit{8\pm2}{\milli\electronvolt}$ (surface defects), $\unit{10\pm2}{\milli\electronvolt}$ (TO-replica) and $\unit{10\pm2}{\milli\electronvolt}$ (M1 mode) for non-radiative recombination processes. To obtain these fits we only took into account $I_{PL}(T)$ for \mbox{$T<\unit{100}{\kelvin}$} where sufficiently intense signal from mode, defect band and TO-replica are available to make a valid comparison. From the similar values of $E_A$ for the cavity mode and TO-replica we conclude that the mode is predominantly excited via the TO-replica, as already deduced from the data shown in \mbox{fig. \ref{figure3} (c)}.

Comparing mode and TO-replica in \mbox{fig. \ref{figure4} (c)}, we see from \mbox{eqn. \ref{eq:6}} that the lower values of \mbox{ln($I_{PL}(T^{0K})/I_{PL}(T) -1)$} from the mode emission can only be explained via the term \mbox{$\Gamma_{Rad}/\Gamma_{NR}^0$}, since both emission peaks exhibit the same activation energy. Hence, the data suggests that the presence of the cavity mode results in a ratio of $\Gamma_{Rad}/\Gamma_{NR}^0$ which is larger than for the TO-replica.
As shown in refs. \cite{Kaniber07,Iwamoto07}, an enhanced PL signal can also be caused by a redistribution of the spatial emission profile via the photonic crystal structure. However, this effect does not increase the radiative recombination rate. Therefore, the results indicate that the enhanced photoluminescence from the cavity mode may be caused by an enhanced internal quantum efficiency $\eta_{int}(T)$ due to a larger radiative carrier recombination rate caused by the Purcell-effect \footnote{An unambiguous prove of Purcell-effect requires time-resolved PL measurements.}. This would explain the observed temperature stability of the mode emission up to room-temperature, in strong contrast to the vanishing background emission for $T\geq150$~K.

In conclusion, we presented a temperature dependent investigation of PL in Si PhC nanocavities. We suggest two mechanisms being responsible for the luminescence of cavity modes, namely phonon-mediated recombination from charge carriers in the electronic band states and recombination from charge carriers trapped in surface defect states. Furthermore, we observed an enhanced internal quantum efficiency in spectral resonance with the cavity mode emission.

We acknowledge financial support from the German Excellence Initiative via the Nanosystems Initiative Munich (NIM), the TUM International Graduate School of Science and Engineering (IGSSE) and the TUM Institute for Advanced Study (IAS).

\bibliography{SiPaper-JJF}

\end{document}